\pdfoutput=1
\documentclass[reprint, amsmath, amssymb, aps, nofootinbib]{revtex4-1}
\usepackage[pdftex]{graphicx}
\usepackage{wasysym}
\usepackage{amsfonts}
\usepackage{ulem}

\newcommand{\COMMENT}[1]{}

\newcommand{\neqa}{\nonumber\end{eqnarray}}

\newcommand{\eqn}[1]{eq.~(\ref{#1})}

\newcommand{\<}{{\langle}}
\renewcommand{\>}{{\rangle}}

\newcommand{\zb}{\bar{z}}

\newcommand{\re}{\relax{\rm I\kern-.18em R}}

\def\su2{{SU(2)}}

\def\[{\left[}
\def\]{\right]}

\def\del{\partial}

\def\to{\rightarrow}

\def\({\left(}
\def\){\right)}
\def\[{\left[}
\def\]{\right]}

\def\<{\langle}
\def\>{\rangle}

\def\i2{\frac{i}{2}}

\def\2F1{\,_2{\rm F}_1}

\usepackage[usenames,dvipsnames]{xcolor}
\usepackage{color}
\usepackage{graphicx}
\usepackage{dcolumn}
\usepackage{bm}
\usepackage{hyperref}

\newcommand{\beq}{\begin{equation}}
\newcommand{\eeq}{\end{equation}}
\newcommand{\beqq}{\begin{equation*}}
\newcommand{\eeqq}{\end{equation*}}
\newcommand\beqa{\begin{eqnarray}}
\newcommand\eeqa{\end{eqnarray}}
\newcommand\beqaa{\begin{eqnarray*}}
\newcommand\eeqaa{\end{eqnarray*}}
\newcommand\bea{\begin{array}}
\newcommand\eea{\end{array}}

\begin{document}

\null\vskip-10pt \hfill
\begin{minipage}[t]{42mm}
SLAC--PUB--16967\\
\end{minipage}
\vspace{0mm}

\title{Gluing Ladders into Fishnets}

\author{Benjamin Basso$^{1}$ and Lance J. Dixon$^{2}$}

\affiliation{
\vspace{5mm}
$^{1}$Laboratoire de physique th\'eorique, D\'epartement de physique de l'ENS, \'Ecole normale sup\'erieure, PSL Research University, Sorbonne Universit\'es, UPMC~Univ.~Paris~06, CNRS, 75005 Paris, France\\
$^{2}$SLAC National Accelerator Laboratory,
Stanford University, Stanford, CA 94309, USA\\
Kavli Institute for Theoretical Physics,
University of California Santa Barbara, CA 93106, USA\\
Institut Philippe Meyer \& Laboratoire de physique th\'eorique,
D\'epartement de physique de l'ENS, \'Ecole normale sup\'erieure,
75005 Paris, France}

\begin{abstract}

\vspace{3mm}

We use integrability at weak coupling to compute
fishnet diagrams for four-point correlation functions
in planar $\phi^4$ theory.  The results are always multi-linear
combinations of ladder integrals, which are in turn built
out of classical polylogarithms. The Steinmann relations provide
a powerful constraint on such linear combinations, leading to 
a natural conjecture for any fishnet diagram as the determinant
of a matrix of ladder integrals.

\end{abstract}

\maketitle

\section{Introduction and Main Result} 

Integrability is a powerful tool for exploring theories such as
planar ${\cal N}=4$ super-Yang-Mills (SYM) theory at finite
coupling~\cite{Beisert2010jr,Basso2013vsa,Gromov2013pga,Basso2015zoa,Fleury2016ykk,Eden2016xvg}.
It can also assist in the computation of individual Feynman diagrams,
in scalar theories
directly~\cite{Zamolodchikov1980mb,Isaev2003tk}, or after suitably
twisting the SYM theory~\cite{Gurdogan2015csr,Caetano2016ydc,Chicherin2017cns},
or, more implicitly, through the ``hexagonalization'' of correlation
functions~\cite{Fleury2016ykk}.

The Steinmann relations~\cite{Steinmann} provide stringent
analytic constraints on multi-particle
scattering amplitudes by forbidding double discontinuities
in overlapping channels. 
They have been applied extensively
in the multi-Regge limit, e.g.~in refs.~\cite{Stapp1982mq,Bartels2008ce}.
Their far-reaching consequences outside of this limit were
recognized more recently.
Combined with the dual conformal symmetry of scattering amplitudes
in the SYM theory, they severely restrict the
types of functions that can appear, making it possible to bootstrap
the six-point amplitude to five
loops~\cite{CaronHuot2016owq} and the (symbol of the) seven-point
amplitude to four loops~\cite{Dixon2016nkn} with very little
additional input.

\begin{figure}[t]
\centering
\def\svgwidth{7cm}
\includegraphics[width=0.27\textwidth]{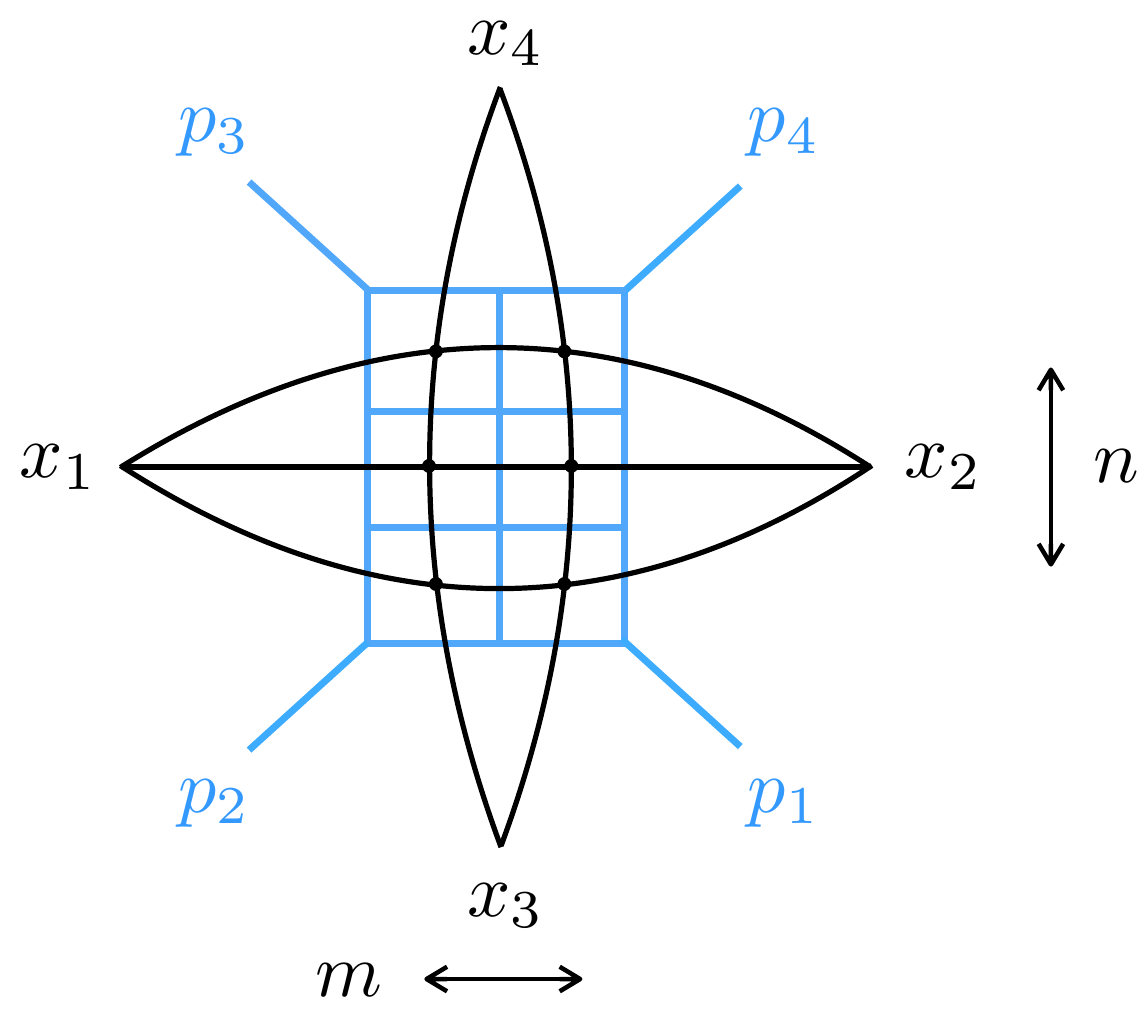}
\vspace{0cm}
\caption{Fishnet diagram in $\phi^4$ theory and its
dual off-shell (color-ordered) scattering amplitude.}
\label{fishnet}
\end{figure}

In this Letter, we combine integrability and the Steinmann relations
in order to find a simple (conjectural) result for the doubly-infinite
class of Feynman graphs depicted in fig.~\ref{fishnet}.
They belong to a broader family of conformal integrals which has
attracted much attention over the
years~\cite{Usyukina1993ch,Broadhurst1993ib,Drummond2006rz,Broadhurst2010ds,%
Drummond2013nda,Eden2016dir,Drummond2012bg,Schnetz2013hqa,Golz2015rea}.
The black lines in the figure provide the position-space interpretation
of the ``fishnet'' diagram, as a contribution to the correlation function 
$G_{m,n}(x_i) = \langle \phi_2^n(x_1) \phi_2^{\dagger n}(x_2)
                      \phi_1^m(x_3) \phi_1^{\dagger m}(x_4) \rangle$, at weak coupling,
$g^2 \equiv \lambda/(4\pi)^2 \ll 1$, 
with $\phi_{1,2}$ two orthogonal complex scalars, $\phi_{1,2}^{\dagger}$
their complex conjugates, and with $\lambda=g_{\rm YM}^2 N_c$ the 't Hooft coupling.

%%%%%%%%%%%
%%%%%%%%%%%

We are only interested in the first planar graph contributing to this
correlator. Given the $R$-charge assignment, all lines must cross each other, as in fig.~\ref{fishnet}
with the scalars' quartic coupling $\lambda/(2\pi)^4 = g^2/\pi^2$
(cf.~the 10-point graph considered in ref.~\cite{CaronHuot2012ab}).
After integrating $\int d^4 x_k$ over each intersection point $x_k$, $k>4$,
and extracting a factor of the disconnected free propagators,
this very first contribution to the correlator reads
\beq
G_{m,n}(x_i) = \frac{g^{2mn}}{(x_{12}^2)^n (x_{34}^2)^m}\times \Phi_{m,n}(u,v) \,,
\label{Gmn}
\eeq
where $x_{ij} = x_{i}-x_{j}$. The two conformal cross ratios are
\beq
u = \frac{x_{14}^2 x_{23}^2}{x_{12}^2 x_{34}^2} \equiv \frac{z\zb}{(1-z)(1-\zb)} \,,
\,\, v = \frac{x_{13}^2 x_{24}^2}{x_{12}^2 x_{34}^2} \equiv \frac{u}{z\zb} \,.
% = \frac{1}{(1-z)(1-\zb)}\, . 
\eeq

Alternatively, we could use the strongly twisted theory considered in
refs.~\cite{Gurdogan2015csr, Caetano2016ydc, Chicherin2017cns}. In that theory,
the gluons and fermions are decoupled, the correlator (\ref{Gmn}) is a
particular instance of the off-shell amplitudes discussed in
ref.~\cite{Chicherin2017cns}, and fig.~\ref{fishnet}
is the only diagram contributing to it.

The blue lines in fig.~\ref{fishnet} indicate a dual-graph, or 
``momentum-space'' (but not Fourier-transformed),
interpretation of the quantity as a contribution
to a scattering amplitude with four external massive momenta,
$p_1=x_{23}$, $p_2=x_{31}$, $p_3=x_{14}$, $p_4=x_{42}$,
and all massless internal lines.
The Steinmann relations~\cite{Steinmann} forbid double discontinuities in
the overlapping channels $(p_1+p_2)^2=x_{12}^2$ and $(p_2+p_3)^2=x_{34}^2$.
The momentum-space interpretation looks like $m$ ladders glued together.
The ladder integrals, corresponding to $m=1$, were computed long
ago~\cite{Usyukina1993ch} in terms of classical polylogarithms.
They also belong to a class of iterated integrals called
single-valued harmonic polylogarithms (SVHPLs)~\cite{BrownSVHPLs}
with weight (number of iterated integrations) equal to $2n$,
where $n$ is the loop number.  The ladder integrals are the
building blocks for the fishnet integrals.

% \noindent\textit{Definition}

We find that $\Phi_{m, n}(u,v)$ can be written, for $m \leq n$, as
\beq
\Phi_{m, n}(u,v)
= \bigg[\frac{(1-z)(1-\zb)}{z-\zb}\bigg]^{m} I_{m, n}(z, \zb)\, ,
\label{PhiandI}
\eeq
where $I_{m, n}$ is an iterated integral (also known as a pure function)
of weight $2mn$.
It is symmetric under $3\leftrightarrow 4$ (equivalently,
$u\leftrightarrow v$, or $z, \zb\leftrightarrow 1/z, 1/\zb$)
and under $z\leftrightarrow \zb$, up to a sign,
\beq
I_{m, n}(1/z, 1/\zb) = I_{m, n}(\zb, z) = (-1)^{m} I_{m, n}(z, \zb)\, .
\eeq
%

% \noindent\textit{Main result}

Our main result is that $I_{m,n}$ is the determinant of an
$m \times m$ matrix,
\beq
I_{m,n} = \textrm{det}\, M, \qquad
M_{ij} = c_{ij} \, L_{n-m-1+i+j} \,.
\label{DetFormula}
\eeq
The matrix elements are $1\times p$ ladder integrals $L_p$
(see \eqn{Ldef}) multiplied by rational numbers,
\beq
c_{ij} =
\begin{cases}
1, & i=j, \\
\prod_{k=j+1}^i p_k(p_k-1),\quad & i> j, \\
[ c_{ji} |_{n\to n+j-i} ]^{-1}, & i<j,
\end{cases}
\label{cij}
\eeq
where $p_k = n-m-1+j+k$.  In the following, we discuss how integrability
and analyticity lead to \eqn{DetFormula}.

%%%%%%%%%%%%%%%%%%%%%%%%%%%%%%%%%%%%%%%%%%%%

\section{Pentagons, hexagons, and all that}

In this section, we present two matrix-model-like
integral representations for the diagram in fig.~\ref{fishnet}, using the 
integrability of planar SYM theory. They correspond to two different ways
of factorizing the fishnet diagram, using the so-called flux-tube
picture~\cite{Alday2007mf,Basso2013vsa}, where the operators are
inserted along the edges of a null Wilson loop, or the more
recent approach proposed to study three-~\cite{Basso2015zoa} and higher-point
functions~\cite{Fleury2016ykk,Eden2016xvg}.\\

%%%%%%%%%%%%
\begin{figure}
%\vspace{5cm}
\centering
\includegraphics[width=0.25\textwidth]{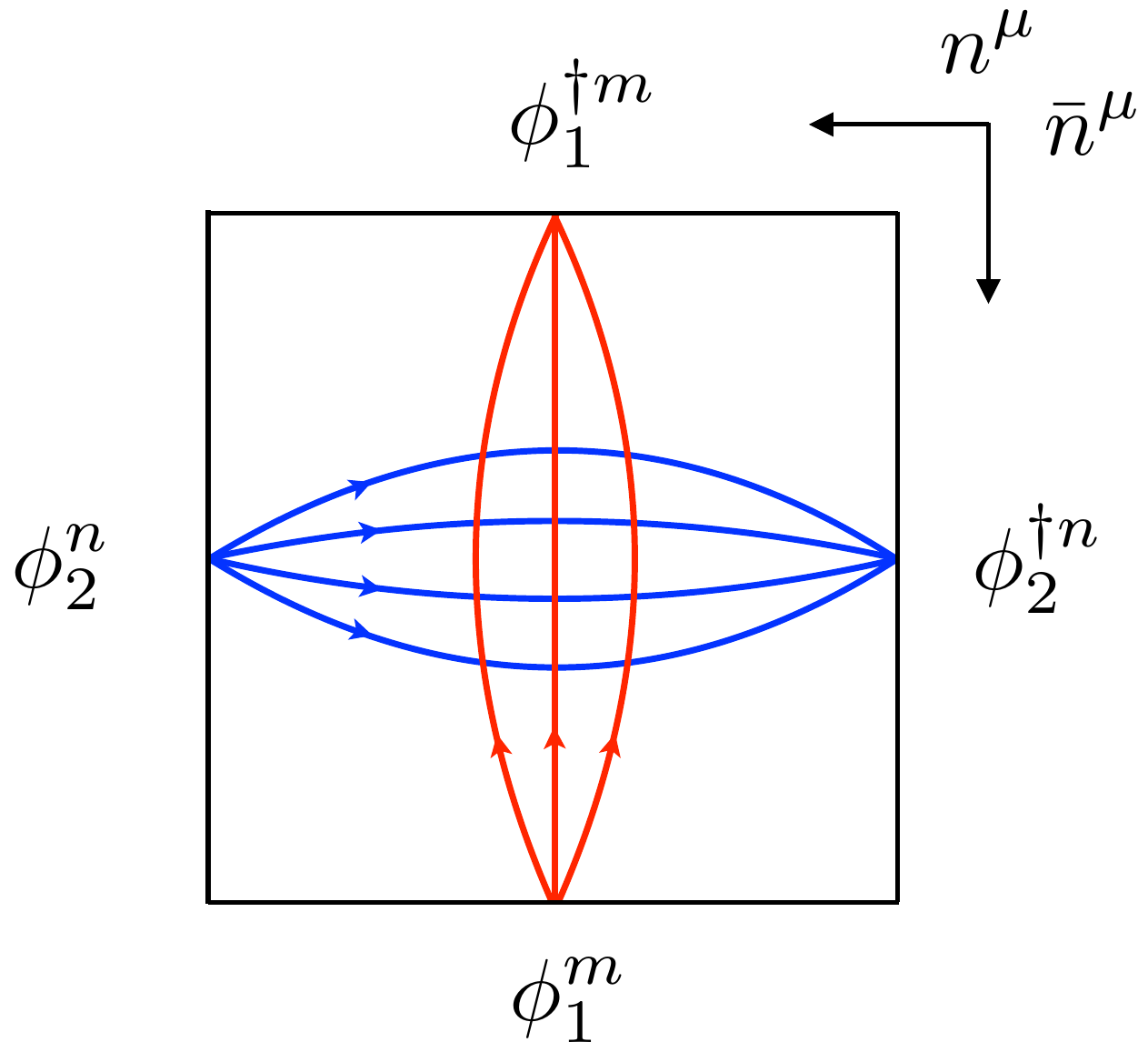}
\vspace{-0cm}
\caption{The correlator can be put inside a null square Wilson loop, with $x_{4}^{\mu}  = n^{\mu}, x^{\mu}_{2} = \bar{n}^{\mu}, \bar{n}x_{3} = -e^{2\sigma_{1}}, nx_{1} = -e^{-2\sigma_{2}}$, and $n^2 = \bar{n}^2 = 0, n\cdot \bar{n} = 1$. Moving $\phi_{1,2}$ along the edges is the same as changing the cross ratios $z = -e^{2\sigma_{1}}$ and $\bar{z} = -e^{-2\sigma_{2}}$.
}\label{picture}
\end{figure}
%%%%%%%%%%%%

\noindent\textit{Flux tube picture.} In the flux tube picture (fig.~\ref{picture}) the two cross ratios map to the positions $\sigma_{1,2}$ of the operators along two light-like directions, $z = -e^{2\sigma_{1}}, \zb = -e^{-2\sigma_{2}}$.
The correlator is viewed as a scattering
of two beams on top of the Gubser-Klebanov-Polyakov~\cite{GKP}
background. The beams are labelled by the scalars' rapidities, $\textbf{u} = \{u_{1}, \ldots , u_{m}\}, \textbf{v} = \{v_{1}, \ldots , v_{n}\}$, which are conjugate to shifts in $\sigma_{1}$ and $\sigma_{2}$, respectively, and are separately conserved throughout the entire process, thanks to integrability. The form factor for the creation and absorption of a beam at the boundary of the square, or equivalently the absolute value of the beam's wave function, can be parametrized in terms of pentagon transitions $P$~\cite{Basso2013vsa}:
\beq\label{overlap}
\mu(\textbf{u}; \sigma_{1}) = \prod_{i=1}^{m} \mu(u_{i})e^{2iu_{i}\sigma_{1}}
\prod_{i<j}^{m}\frac{1}{P(u_{i}|u_{j})P(u_{j}|u_{i})}\, ,
\eeq
with $\mu(u) = \pi \, \textrm{sech}{(\pi u)}$ and $P(u|v) = \Gamma(iu-iv)/\Gamma(\tfrac{1}{2}+iu)\Gamma(\tfrac{1}{2}-iv)$.
%%
%\beq
%\mu(u) = \frac{\pi}{\cosh{(\pi u)}}\, ,
%\quad P(u|v)
%= \frac{\Gamma(iu-iv)}{\Gamma(\tfrac{1}{2}+iu)\Gamma(\tfrac{1}{2}-iv)} \, .
%\eeq
%%
Integrating~(\ref{overlap}) over the rapidities gives back the free
propagator for $m$ scalar fields inserted along the null direction,
\beq\label{ds}
d(\sigma_{1})^{-m} = \bigg[\frac{1}{e^{\sigma_{1}}+e^{-\sigma_{1}}}\bigg]^{m} = \int \frac{d\textbf{u}}{m!}\, \mu(\textbf{u}; \sigma_{1})\, ,
\eeq
with $d\textbf{u} = \prod_{i}du_{i}/(2\pi)$ and
$\mathcal{h}\sqrt{-z}\phi(x_3)\, \phi^{\dagger}(x_4)\mathcal{i} = \sqrt{-z}/(1-z) = 1/d(\sigma_1)$.
Eq.~(\ref{ds}) is also the spin-chain scalar product in the so-called separated variables \cite{Derkachov2002tf}. The same expression with $u_{i} \rightarrow v_{i}$, $m\to n$, $\sigma_{1}\to\sigma_{2}$ describes the second beam.

An essential property of flux-tube scattering is that it is diffractionless and fully factorized. Hence, the $m\times n$ grid in the diagram can be immediately taken into account by inserting $\prod_{i=1}^{m}\prod_{j=1}^{n} S_{\star}(u_{i}, v_{j})$,
%%
%\beq
%\prod_{i=1}^{m}\prod_{j=1}^{n} S_{\star}(u_{i}, v_{j})\, ,
%\eeq
%%
%in rapidity space,
where $S_{\star}(u, v)$ is the transmission part of the mirror two-body $S$-matrix \cite{Basso2013pxa},
\beq
S_{\star}(u, v) = \frac{\pi g^2 \sinh{(\pi(u-v))}}{(u-v)\cosh{(\pi u)}\cosh{(\pi v)}}\, .
\eeq
The overall process is of order $O(g^{2mn})$, in agreement with the corresponding Feynman diagram.

Assembling all factors together, and dropping the powers of the coupling, we obtain the flux tube representation
\beq\label{flux-tube}
\begin{aligned}
&\frac{\Phi_{m,n}}{d_{1}^{m}d_{2}^{n}} = \int \frac{d\textbf{u}d\textbf{v}}{m!n!}
\prod_{i}^{m}\mu(u_{i})^{m+n} e^{2iu_{i}\sigma_{1}}
\prod_{i}^{n}\mu(v_{i})^{m+n}e^{2iv_{i}\sigma_{2}} \\
& \qquad \qquad \times \prod_{i< j}^{m}\Delta(u_{i}, u_{j})
\prod_{i,j}^{m,n}\tilde{\Delta}(u_{i}, v_{j})\prod_{i< j}^{n}\Delta(v_{i}, v_{j}),
\end{aligned}
\eeq
where $d\textbf{v} = \prod_{i}dv_{i}/(2\pi)$,
%
%$d\textbf{u}$ was defined earlier, $d\textbf{v}$ is defined similarly,
%
$d_{i} = d(\sigma_{i})$,
\beq
\tilde{\Delta}(u, v) = \frac{\sinh{(\pi(u-v))}}{\pi (u-v)}
= \frac{\Delta(u, v)}{(u-v)^2} \, ,
\eeq
%
%\beq
%\begin{aligned}
%&\Delta(u, v) = \frac{1}{\pi} (u-v)\sinh{(\pi(u-v))}\, , \\
%&\tilde{\Delta}(u, v) = \frac{\sinh{(\pi(u-v))}}{\pi (u-v)} \, .
%\end{aligned}
%\eeq
and dividing by the disconnected propagators matches the
normalization~(\ref{PhiandI}).  A similar integral has been used
to study 2-to-2 fermion flux-tube scattering~\cite{Basso2014koa}.\\

\begin{figure}[t]
\vspace{1cm}
\centering
\includegraphics[width=0.45\textwidth]{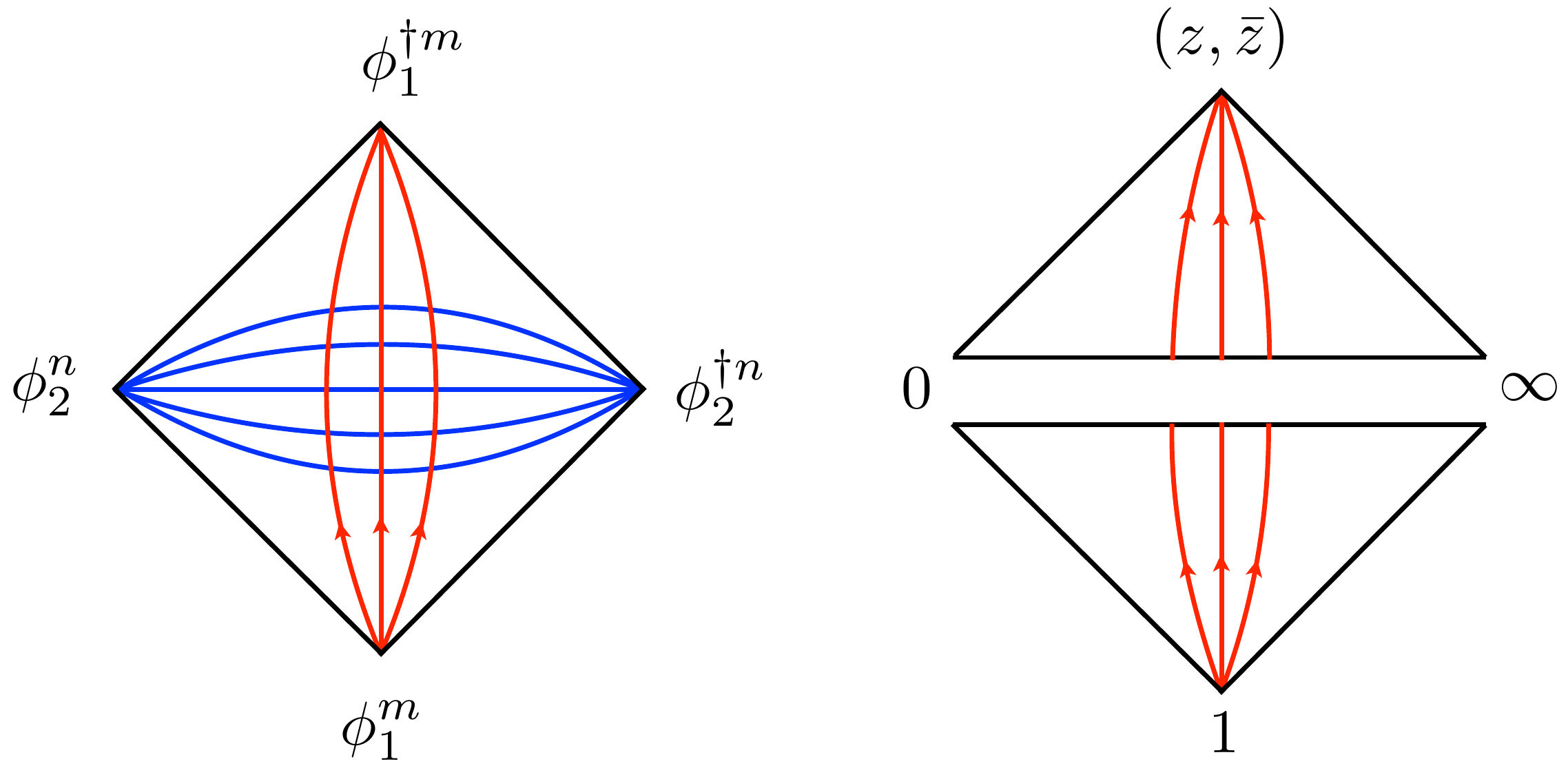}
\vspace{-0cm}
\caption{We can decompose the correlator using triangles (also known as hexagons). The red beam is made of $m$ magnons, produced on the bottom triangle and absorbed on the top one. The correlator is the scalar product between the two wave functions.
}\label{trianglepicture}
\end{figure}

\noindent\textit{BMN picture.} An alternative representation for the same
correlator comes from the Berenstein-Maldacena-Nastase (BMN)~\cite{BMN} picture. In this picture, one beam, $\phi_{2}^{n}$, describes a reference
state, the BMN vacuum, while the other, $\phi_{1}^{m}$, is viewed as a collection of $m$ magnons propagating through it. The latter are not the familiar magnons describing spin waves on top of the (ferromagnetic) vacuum, but some ``mirror" versions of them, mapping to insertions along the direction $(x_{1}, x_{2}) = (0, \infty)$ of the reference beam, see fig.~\ref{trianglepicture}. Each magnon $\phi_{1}$ is further decomposed into partial waves with respect to dilatation and rotation, $z = \rho e^{i\phi}$; each carries a rapidity $u\in \mathbb{R}$ and bound state index $a\in \mathbb{Z}$ conjugate to these symmetries. The planar correlator is cut halfway by the vacuum into two triangles,
which are naturally associated with three-point functions.
The amplitudes for production and absorption of $m$ magnons on the
two triangles can be obtained in terms of the so-called hexagon form
factors~\cite{Basso2015zoa,Basso2015eqa}. The next crucial ingredient is
the rule for rotating each partial wave from a triangle ending on the
reference points (0, 1, $\infty$) to the reference points
$(0, z, \infty)$~\cite{Fleury2016ykk}.
Combining the two yields the wave-function overlap
\beq\label{wf}
\begin{aligned}
&\mu_{\textbf{a}}(\textbf{u};  z) = \frac{|z|^m}{(z-\bar{z})^m}\prod_{i=1}^{m}z^{-iu_{i}+\tfrac{a_{i}}{2}}\bar{z}^{-iu_{i}-\tfrac{a_{i}}{2}}\mu_{a_{i}}(u_{i})\prod_{i<j}^{m}p_{ij}\, ,
%& \qquad \qquad \times \prod_{i<j}^{m} p_{a_{i}a_{j}}(u_{i}, u_{j})\, ,
\end{aligned}
\eeq
with $(\textbf{u}, \textbf{a}) =\{(u_{1}, a_{1}), \ldots\}$, $\mu_{a}(u) = a g^2/(u^2+\tfrac{a^2}{4})^2$, $p_{ij} = p_{a_{i}a_{j}}(u_{i}, u_{j})$, $p_{ab}(u, v) = \mu_{a}(u)\Delta_{ab}(u, v)\mu_{b}(v)/(ab)$, and $\Delta_{ab}(u, v)$ as defined in eq.~(\ref{Delta}) below.

\begin{figure}
%\vspace{5cm}
\centering
\includegraphics[width=0.34\textwidth]{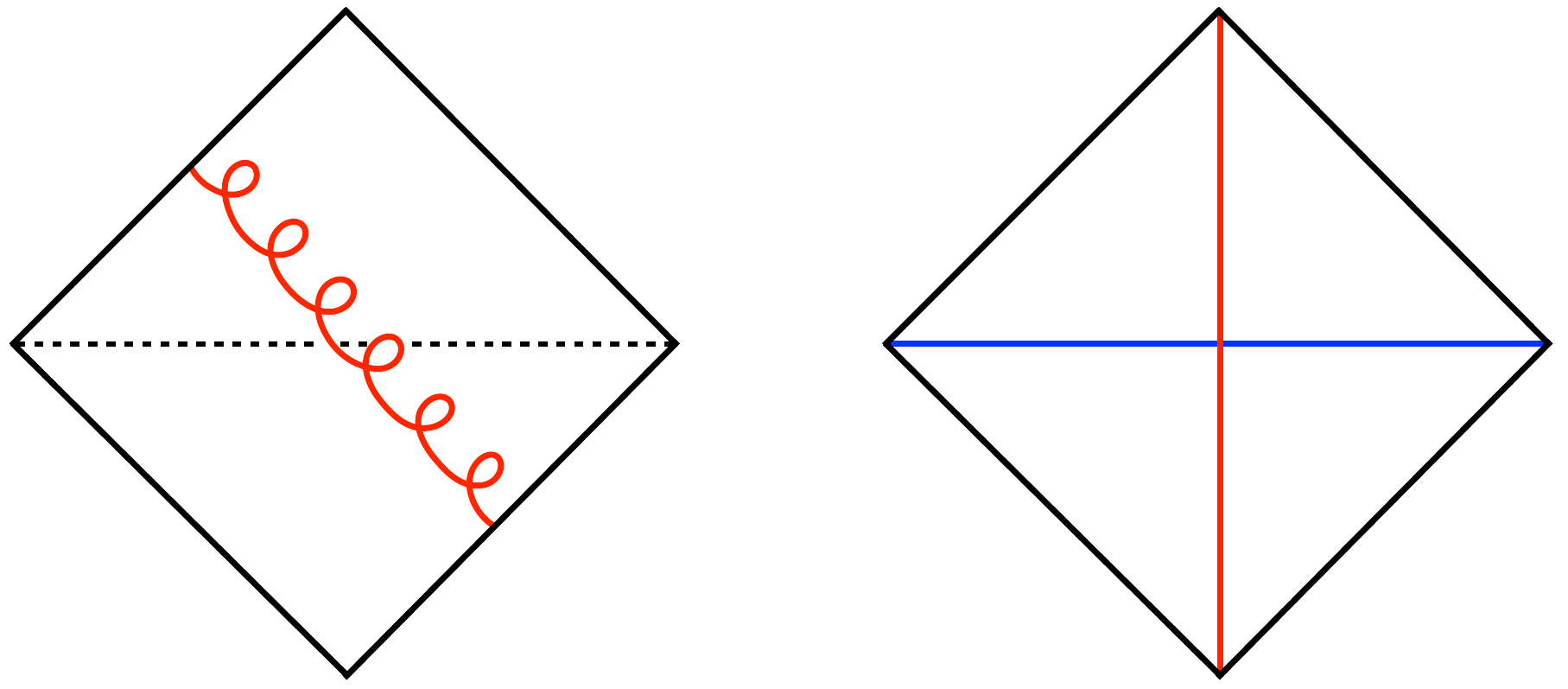}
\vspace{-0cm}
\caption{By supersymmetry, the measure $\mu_{a}(u)$ describes both the one-loop gluon diagram, on the left, and the one-loop scalar cross diagram, on the right. To get a free propagator, we must deconvolute the scalar diagram, by acting with the box operator $\Box /g^2 = -z\bar{z}\partial_{z}\partial_{\zb}/g^2$ or, equivalently, by introducing the form factor $(u^2+a^2/4)/g^2$ in rapidity space. A similar rule was used~\cite{Basso2013vsa,Belitsky2014sla,Basso2015rta} to transpose between MHV and NMHV amplitudes, in the flux tube picture. In general, the conversion is achieved through inclusion of a factor $((u^2+a^2/4)/g^2)^{m}$, per excitation, where $m$ is the ``NMHV" degree, or number of scalars at the cusp. This is readily seen to correct the mismatch between $\ell$ and $n$. 
}\label{picture2}
\end{figure}

Finally, the ``scattering'' between the magnons and the vacuum results in a
factor $(g^2/(u^2+a^2/4))^{\ell}$ per magnon, where $\ell$ is the
so-called bridge length. Naively, $\ell = n$, since there are $n$
vacuum lines to cross. In fact $m$ of these lines have been pulled out
and included in the wave function (\ref{wf}), as shown in 
fig.~\ref{picture2} for $m=1$. This subtlety of the cutting explains why
the wave function (\ref{wf}) is suppressed by $2m^2$ powers of the coupling
and why the bridge length is $\ell = n-m$.  For $n=0$ ($\ell = -m$),
the overlap gives back the tree result, upon integration, 
\beq\label{free}
\sum_{\textbf{a}\in \mathbb{Z}^{m}}\int \frac{d\textbf{u}}{m!} \, \mu_{\textbf{a}}(\textbf{u};  z) \prod_{i}^{m}\frac{(u_{i}^2+\tfrac{a_{i}^2}{4})^{m}}{g^{2m}} = \frac{|z|^m}{|1-z|^{2m}}\, ,
\eeq
with $|z|/|1-z|^2 = \sqrt{x^2_{3}x^2_{4}}/x_{34}^2$ the scalar propagator in the conformal frame of fig.~\ref{trianglepicture} (with the numerator absorbing the weights of the field).

Putting everything together, and normalizing by the disconnected correlator, eq.~(\ref{free}), leads to an integral for the pure function directly,
\beq\label{hexagon}
\begin{aligned}
&I_{m, n} = \sum_{\textbf{a}}\int \frac{d\textbf{u}}{m!}
\prod_{i=1}^{m} \frac{a_{i}z^{-iu_{i}+a_{i}/2}\zb^{-iu_{i}-a_{i}/2}}{(u_{i}^2+a_{i}^2/4)^{m+n}}\prod_{i<j}^{m}\Delta_{ij}\, ,
%&\qquad \qquad \qquad \times \prod_{i<j}^{m}\Delta_{a_{i}a_{j}}(u_{i}, u_{j})\, ,
\end{aligned}
\eeq
with $\Delta_{ij} = \Delta_{a_{i}a_{j}}(u_{i}, u_{j})$ and
\beq\label{Delta}
\Delta_{ab}(u, v) = \biggl[ (u-v)^2+\frac{(a-b)^2}{4} \biggr]
                   \biggl[ (u-v)^2+\frac{(a+b)^2}{4} \biggr] \, .
\eeq
For $m=1$, this is the formula for the one magnon contribution to the four point function of chiral primary operators in planar $\mathcal{N}=4$ SYM~\cite{Fleury2016ykk}.\\

\noindent\textit{Analysis.} The flux tube and BMN matrix integrals provide two formulae for the fishnet diagram which are equivalent, in principle. In practice, it is much easier to evaluate the latter. There are far fewer residues and the answer appears in closed form almost immediately; the final sum over the bound state labels is always expressible in terms of classical polylogarithms. The infinite series of ladder integrals ($m=1$) was easily reproduced in this manner~\cite{Fleury2016ykk}. Thanks to the polynomial nature of the magnon interaction, eq.~(\ref{Delta}), the fishnet diagrams are equally straightforward for reasonable values of $m,n$.  We derived the result (\ref{DetFormula}) through $m, n = 1, \ldots, 4$. We double-checked the answer against the flux-tube predictions for the few lowest residues when $m=2$.  The main structural property, embodied in \eqn{DetFormula}, is that the fishnet diagrams are sums of products of $m$ ladder integrals. This observation is the seed for the Steinmann bootstrap program.

%%%%%%%%%%%%%%%%
%%%%%%%%%%%%%

%%%%%%%%%%%%%%%%%%%%%%%%%%%%%%%%%%%%%%

\section{Ladders, Steinmann, and all that}

The ladder function $L_p$ is defined for $p>0$
by~\cite{Usyukina1993ch,Broadhurst2010ds}
\beq
L_p = \sum_{j=p}^{2p} \frac{j!\ [-\ln(z\zb)]^{2p-j}}{p!(j-p)!(2p-j)!} 
[ {\rm Li}_j(z) - {\rm Li}_j(\zb) ],
\label{Ldef}
\eeq
and the tree-level value is $L_0 = (z-\zb)/[(1-z)(1-\zb)]$.
%
%\beq
%L_0 = \frac{z-\zb}{(1-z)(1-\zb)} \, .
%\label{L0def}
%\eeq
%
In the neighborhood of the origin in $z$, the polylogarithms
${\rm Li}_j(z)$ are analytic, and $L_p$ is manifestly single-valued,
a real analytic funtion of $z$.
That is, $L_p$ has no branch cuts under rotating $z\to ze^{2\pi i}$, 
$\zb\to \zb e^{-2\pi i}$. It does have (multiple) discontinuities
in $z\zb$: under $z\to ze^{\pi i}$, $\zb\to \zb e^{\pi i}$, the logarithm
shifts by $\ln(z\zb)\to\ln(z\zb)+2\pi i$.

What is not so obvious from the representation~(\ref{Ldef}) is
that $L_p$ is also single-valued in $z$ around $z=1$. In fact
it lies in the class of SVHPLs ${\cal L}_{\vec{w}}$~\cite{BrownSVHPLs},
\beq
L_p = (-1)^p \Bigl[ {\cal L}_{0,\ldots,0,1,0,0,\ldots,0}
                  - {\cal L}_{0,\ldots,0,0,1,0,\ldots,0} \Bigr] \,,
\label{SVHPLform}
\eeq
where there are $p-1$ ($p$) 0's before the 1 in the first (second) term,
and $2p$ entries in all.

Now $L_p$ does have a discontinuity
in $(1-z)(1-\zb)$: under $(1-z)\to (1-z)e^{\pi i}$,
$(1-\zb)\to (1-\zb) e^{\pi i}$,
\beqa
L_p &\to& L_p + {\rm Disc}_1 L_p  \,, \label{discLp}\\
{\rm Disc}_1 L_p &=&
2\pi i  \frac{(-1)^p}{p!(p-1)!} \, \ln(z/\zb) \, ( \ln z \ln\zb )^{p-1} \,.
\label{Ldisc}
\eeqa
This discontinuity is compatible with the differential equation
obeyed by $L_p$~\cite{Drummond2006rz}:  
\beqa
z\zb \del_z \del_{\zb} L_p &=& - L_{p-1} \,, \label{diff_eq}\\
z\zb \del_z \del_{\zb} {\rm Disc}_1 L_p &=& - {\rm Disc}_1 L_{p-1} \,.
\label{diff_eq_disc}
\eeqa
Crucially, \eqn{discLp} is only a {\it single} discontinuity,
due to the Steinmann relations~\cite{Steinmann}
for the momentum-space interpretation of the integral.

The Steinmann relations forbid a double discontinuity in the
overlapping $s$ and $t$ channels of the four-point
amplitude for massive scattering:
\beq
{\rm Disc}_{s} {\rm Disc}_{t} {\cal A}_4 = 0,
\label{SteinmannA}
\eeq
where $s=x_{12}^2$, $t=x_{34}^2$.
Conformal invariance places $s$ and $t$ both in the denominator of $u$ and $v$,
so the discontinuities now take place in the {\it common}
variable $(1-z)(1-\zb)$ at $z=1$, and \eqn{SteinmannA} becomes
\beq
{\rm Disc}_{1} {\rm Disc}_{1} {\cal A}_4 = 0.
\label{SteinmannB}
\eeq
This equation holds for any conformally-invariant
Feynman integral with the same kinematics, such as $I_{m,n}$
or $L_p$.\footnote{%
The position-space interpretation of the single allowed discontinuity
is in terms of an extremal process where a twist-$2n$ operator
$\sim \phi_{1}^{m}\Box^{\ell}\phi_{1}^{\dagger m}$, with $\ell=n-m$,
is exchanged between the two beams. (Schematically, the $n-m$ boxes
in the operator remove $\ell = n-m$ propagators in the bridge, giving rise
to an effective bridge with $\ell' = 0$, which is characteristic of extremal
processes.) Its contribution to the correlator is power-suppressed,
appearing at order $|1-z|^{2n}$ in the expansion of $\Phi_{m,n}$ around
$z=\bar{z} = 1$, but it is logarithmically enhanced because of mixing
between single- and double-trace operators. In the planar limit,
one typically expects a single logarithm from double-trace mixing.}
(Many conformal integrals, e.g.~those considered in ref.~\cite{Drummond2012bg},
don't have a scattering interpretation, so the Steinmann relations don't
apply to them.)

Generic products of ladder integrals do {\it not} obey the Steinmann relations,
because the single discontinuities in $(1-z)(1-\zb)$
multiply together to form multiple discontinuities.
In special combinations, the multiple discontinuities cancel.
For example, in the linear combination
\beq
L_{n-1} L_{n+1} + r (L_n)^2 \,,
\eeq
the $z$ dependence of the double discontinuity in each term
is precisely the same,
and the respective normalization factors are $[(n+1)!n!(n-1)!(n-2)!]^{-1}$
and $[(n!)^2 ((n-1)!)^2]^{-1}$.  If $r=-(n-1)/(n+1)$,
then the double discontinuity cancels between the two terms.
This value of $r$ agrees with the direct computation and
gives the $m=2$ result $I_{2,n}$ in \eqn{DetFormula} ($r=-c_{12}c_{21}$).

For $m=n=2$, the integral $I_{2,2} = L_1 L_3 - \tfrac{1}{3} (L_2)^2$
can be evaluated using~\eqn{SVHPLform}.  Converting
the ${\cal L}_{\vec{w}}$ functions to a linearized form with
shuffle identities, and using the compressed notation
of ref.~\cite{Eden2016dir}, we obtain
\beqa
I_{2,2} &=& 4 [ - {\cal L}_{3,5} + {\cal L}_{5,3} + {\cal L}_{2,5,0} 
  - {\cal L}_{4,3,0} - {\cal L}_{1,5,0,0} \nonumber\\
&&\hskip0.2cm\null
 + {\cal L}_{3,3,0,0} - {\cal L}_{2,3,0,0,0} + {\cal L}_{1,3,0,0,0,0} ] \,,
\label{compareES}
\eeqa
a form which agrees with ref.~\cite{Eden2016dir}.

The cancellation of multiple discontinuities becomes a very stringent
requirement as the number of ladders increases.  A particular term
always appears with unit coefficient in the $m\times n$ fishnet result:
$L_{n-m+1} L_{n-m+3} \ldots L_{n+m-1}$.
For the square fishnet with $m=n$, we write all combinations of $m$ ladders
$L_{p_i}$ with weight $2mn=2m^2$, whose maximum index is $p_{\rm max}=2m-1$.
Through $m=n=9$, there is a unique solution to the Steinmann constraints, with
$1,2,5,16,58,231,\ldots$ terms for $m=1,2,3,\ldots$.  This
sequence is the number of monomials in the expansion of the determinant
of the $m\times m$ Hankel matrix $A_{ij}$ with elements
$a_{i+j}$~\cite{OEISA019448} --- a strong clue to
the final formula~(\ref{DetFormula}).

We promote the $m\times m$ solution to an $m\times n$ ansatz by
shifting the arguments of all $L_p$'s in the $m\times m$ solution
upward by $(n-m)$, increasing the weight from $2m^2$ to $2mn$, and
inserting arbitrary functions of $n$ as coefficients of these monomials.
That is, we assume that there are the same number of terms in the
$m\times n$ result as in the $m\times m$ one, and we assume the unit
coefficient in front of $L_{n-m+1} L_{n-m+3} \ldots L_{n+m-1}$.
Through at least $m=8$, the Steinmann constraints have a
unique solution, \eqn{DetFormula}.

We now show that \eqn{DetFormula} solves the Steinmann
constraint~(\ref{SteinmannB}) for any $m,n$.
Notice that the coefficients $c_{ij}$ in \eqn{cij}
and the ladder discontinuities obey very similar relations,
moving along a column of the matrix $M_{ij}$:
\beqa
c_{i+1,j} &=& p(p+1) \, c_{ij} \,,
\label{cRelation}\\
{\rm Disc}_{1} L_{p+1} &=& - \frac{\ln z\ln\zb}{p(p+1)} \, {\rm Disc}_{1} L_{p} \,,
\label{DiscRelation}
\eeqa
where $p=n-m-1+i+j$ is the index for the ladder
$L_p$ that multiplies $c_{ij}$ in $M_{ij}$.
Thus, under \eqn{discLp} every column in $M$ shifts by an amount
proportional to the transpose of the vector
\beq
( 1, -\ln z\ln\zb, [-\ln z\ln\zb]^2, \ldots, [-\ln z\ln\zb]^m ) \,.
\label{commonshift}
\eeq
The double discontinuity in $\det M$ can be computed by summing over
all possible pairs of shifted columns; the determinant
of each such term vanishes because the two columns are proportional.
Therefore the double discontinuity --- and similarly, all higher
discontinuities --- vanish in $I_{m,n}$.
Only the single discontinuity survives.

The Steinmann relations are homogeneous and don't fix the result's overall
normalization.  We check the normalization recursively
in $m$ by observing that \eqn{DetFormula}, although intended to be used for
$n\geq m$, also holds for $n=m-1$, with $I_{m,m-1} = L_0 \, I_{m-1,m}$.
The factor of $L_0$ cancels one inverse factor in \eqn{PhiandI}
for $\Phi_{m,m-1}$, so that $\Phi_{m,m-1} = \Phi_{m-1,m}$ as required
for self-consistency.

%%%%%%%%%%%%%%%%%%%%%%%%%%%

\section{Conclusions}

In this letter we presented a well-motivated conjecture for
conformal four-point fishnet diagrams in terms of ladder integrals.
One may be able to test our conjecture further, by computing the two
integrability-based formulae exactly, for any $m,n$, and proving their 
equivalence to \eqn{DetFormula}. Determinantal representations for the 
integrands, like the one studied in ref.~\cite{Jiang2016ulr},
might enable their exact integration. The conversion between the flux-tube
and BMN pictures might help to find representations of more general
correlators in the separated variables. It might also shed light on
the hidden simplicity of general flux tube integrals,
and bridge the gap to the amplitude bootstrap
program~\cite{CaronHuot2016owq,Dixon2016nkn,Dixon2011pw,Drummond2014ffa}.

One could apply similar techniques to related diagrams,
at the four- and higher-point level. Some alterations of fishnet graphs, either in the bulk or at the boundary, might admit a natural interpretation in the integrability set-up, like ones explored~\cite{Caetano2016ydc} for two-point functions.  Some might echo the magic identities relating many conformal four-point integrals to one another and to the ladder integrals~\cite{Drummond2006rz}. 
When these integrals are ``glued'' together in various ways, are multi-linear
combinations of ladder integrals still obtained?  We expect the combination of
integrability and analyticity to answer that question and lead to 
many more powerful results in the future.
\\
\\
{\it Acknowledgments:} We thank J.~Bourjaily, J.~Caetano, S.~Caron-Huot,
J.~Drummond, T.~Fleury, \"{O}.~G\"{u}rdo\u{g}an,
H.~Johansson, V.~Kazakov, S.~Komatsu, A.~Sever and P.~Vieira for enlightening
discussions and comments on the manuscript. We also thank D.~Zhong for help with the pictures. This research was supported by the US Department of Energy under
contract DE--AC02--76SF00515 and by the National Science Foundation under 
Grant No.~NSF PHY-1125915.
LD is grateful to LPTENS and the Institut de Physique Th\'eorique 
Philippe Meyer for hospitality during this work's initiation,
and to the Kavli Institute for Theoretical Physics and the Simons
Foundation for hospitality during its completion.

\end{document}